\documentstyle[hx37_latex,overcite]{article}

\def\be{\begin{equation}}
\def\ee{\end{equation}}
\def\bea{\begin{eqnarray}}
\def\eea{\end{eqnarray}}
\def\ang{\thinspace{\rm \AA}}
\def\ha{{H$\alpha$}}

\def\msun{{$M_{\odot}$}}

\def\ten#1{10^{#1}}
\def\la{{Ly$\alpha$}}
\def\civ{{\rm C}\thinspace{IV}}
\def\oii{[{\rm O}\thinspace{II}]}
\def\etal{{\it et al.\/}}
\def\qnought{{$q_0$}}
\def\erg{${\rm erg}\ {\rm s}^{-1}$}

\begin{document}

\title{GALAXIES AT Z $\bf>4.5$}

\author{ESTHER M. HU}

\address{Institute for Astronomy, University of Hawaii, 2680 Woodlawn Dr.,\\
Honolulu, HI 96822, USA}

\author{RICHARD G. MCMAHON}

\address{Institute of Astronomy, Madingley Road, Cambridge CB3 0HA, UK}

\author{EIICHI EGAMI}

\address{Max-Planck-Institut f\"{u}r extraterrestrische Physik, Postfach 1603,\\
85740 Garching, GERMANY}


\maketitle\abstracts{
We present the results of ground-based and HST imaging studies
targeted on $z>4.5$ quasar fields.  High-redshift galaxies identified
in deep narrow-band \la\ images of the fields surrounding the quasars
BR1202--0725 ($z=4.694$) and BR2237--0607 ($z=4.558$) have been
confirmed with follow-up spectroscopy using the LRIS spectrograph on
the Keck 10m telescope.  These high-redshift galaxies are typically
only moderately luminous objects and their sub-$L^{\ast}$ magnitudes
imply star formation rates of only a few solar masses per year.  For
BR1202--0725 tip-tilt imaging in \la\ at the UH 2.2m telescope and 
deep narrow-band IR imaging in \oii\ at CFHT reveals a complex and
structured system which may be merging to produce the host galaxy of
the quasar.  For BR2237--0607 the \la-emitting galaxies show the
strong \la-emission and weak continuum expected for objects before
substantial chemical enrichment and consequent dust formation and
extinction has occurred.  These objects appear to be galaxies in
their first outburst of star formation.}

\section{Introduction}
It is clear that some form of activity took place prior to $z\sim5$
to generate the ionization of the intergalactic medium\cite{rees}
and also that some quasar absorption-line systems have small amounts
of metals in place\cite{lu} by $z=4-5$, indicating that some
galaxy formation has initiated at or prior to these redshifts.  These
early epochs may be our best chance to see galaxies in the first
stages prior to the formation of substantial dust and metals, and
hence may be the best place for \la\ emission-line searches.  While
earlier \la\ searches were unsuccessful\cite{pritchet,tdt95},
improved sensitivities -- particularly with the Keck 10m telescope --
are now enabling us to detect \la\ emitters at $z\sim4.5-5$, at least
around the known $z>4.5$ quasars, and there have been a flurry of
recent results\cite{hu96a,fontana96, petit96,hu96b}.  Parallel
searches for color-selected objects are also taking place, tuned to
these redshift ranges, which are designed to turn up more evolved
objects\cite{egami,fontana96,dodorico}.

We discuss here several high-redshift galaxies which have been
identified in deep narrow-band \la\ imaging of the fields surrounding
the quasars BR1202--0725 and BR2237--0607, and studied with followup
spectroscopy using the LRIS spectrograph on Keck.  These objects
appear to follow very closely our expectations for early galaxies
although, consistent with objects at lower redshifts, they are not
particularly luminous.

\section{The companions to BR1202--0725} 
Both BR1202--0725 and BR2237--0607 were identified in the APM quasar
survey\cite{apm,lsl96a} and show $z>4$ damped \la\ (DLA) systems 
in their spectra\cite{apmdla}.  The damped system at
$z=4.383$ [refs.~\citen{lu,apmdla}] seen in BR1202--0725 is the highest
redshift DLA known.
Narrow-band imaging\cite{hu96a} by Hu, McMahon, and Egami at the 
wavelength of redshifted \la\ for BR1202--0725
identified emission which coincides with the continuum light from a faint
galaxy seen in Hubble Space Telescope WFPC2 images 2.6$''$ NW of
the quasar, and pointing radially towards it.  This object was
also identified as a high-redshift galaxy by Fontana \etal\
using color-break techniques\cite{fontana96}  and by Petitjean \etal\
using integral field spectroscopy\cite{petit96}.
From high-resolution
tip-tilt imaging (FWHM $0.45''$) of this system at the UH 2.2m telescope
in the \la\ line we have now identified a second faint \la\ emission 
component coincident
with a fainter galaxy 3$''$ SW of the quasar, also pointing
radially towards the quasar in the $HST$ image (Fig.~\ref{fig:br1202}).  
In these higher resolution images it is possible to see that the
peak of the \la\ emission coincides with the two continuum structures
seen in the $HST$ images, but that there is also a diffuse component
extending out towards the illuminating quasar.  The second emission
system near the quasar at $z=4.7$ is confirmed in multi-slit LRIS spectra
taken on these objects with the Keck 10m telescope.  \civ\ emission is not
detected in these spectra.  However, \oii\ emission is also seen in these
objects in narrow-band IR filter imaging at CFHT made using the UH $1024^2$
IR camera, QUIRC.  The morphology of these nearby systems suggests that
we may be seeing here the merging of these subsystems and possibly witnessing
the formation of the quasar's host galaxy. Near
these emission-line structures there is excess emission at millimeter
and submillimeter wavelengths\cite{rgm94,isaak} and most recently, 
molecular CO transitions have been detected in close proximity to
the quasar\cite{br1202_co}.

\section{Field Galaxies around BR2237--0607 --- Early Star Formation?}
For BR2237--0607 based on deep narrow-band \la\ imaging followed by
spectroscopy using LRIS at Keck, we identify\cite{hu96b} two
\la\ emitting galaxies at $z=4.55$.  In contrast to the BR1202--0725
emission-line objects, these galaxies have separations $> 100''$
($\sim 700$ kpc) from the quasar, which is unlikely to have a
significant role in exciting them.  Observed equivalent widths for
the \la\ line exceed $\sim$1000\ang, in each case.  While one of
these objects is quite compact, and has no detectable continuum in a
1-hr $I$-band exposure at Keck, and thus might be an AGN, the second
has a diffuse extended structure in both continuum and emission, and
is most likely a star-forming galaxy.  Line fluxes are $\sim5\times
10^{-17}$ ergs cm$^{-2}$ s$^{-1}$.  These objects may represent the
very earliest stages of galaxy formation prior to the formation of a
significant stellar population and the destruction of the \la\ line
by dust.

\section{Discussion}
The objects in the BR2237--0607 field are currently the only galaxies,
well separated from quasars, that are known at these high redshifts, and
their properties are therefore of considerable interest if only in targeting
future searches. The rest frame equivalent widths of the systems are around 
$>130$\ang\ and $>240$\ang, which are marginally consistent with stellar
excitation for an initial mass function dominated by massive
stars\cite{charl93}.  The \la\ may arise from a combination of internal and 
external ionization.  If the observed emission were primarily due to stars, 
and there were no internal scattering and extinction, then the luminosity of
$3\times\ten{42}$ $h^{-2}$ \erg\ (\qnought=0.5) would correspond to a star
formation rate in solar masses (\msun) per year of $\sim3\ h^{-2}$
\msun\ yr$^{-1}$, where we use Kennicutt's (ref.~\citen{kenn83}) relation
between \ha\ luminosity and star formation rate (SFR) of SFR = $L($\ha)
$\times\ 8.9\ \times\ten{-42}$ erg s$^{-1}$ \msun\ yr$^{-1}$, and assume a
ratio of \la\ to \ha\ (8.7) that applies for Case B
recombination\cite{brock}.  Given the Hubble time at this redshift of
$7\times\ten{8}\ h^{-1}$ yr (\qnought=0.5) the integrated amount of star
formation is small compared to that of a `normal' galaxy with
$6\times\ten{10}\ h^{-1}$ \msun\ of stars (a so-called $L^*$ galaxy).

Because of the targeted nature of the search it is hard to estimate from
the present data whether such objects may be common in the general field or
whether they are preferentially found around quasars.  Observations,
currently in progress, of additional quasars and blank field regions
should answer this question.

\begin{figure}
\raggedbottom
\addtolength{\topskip}{0pt plus .0001fil}
\renewcommand{\newpage}{\par \break}
\begin{center}
\
\caption{The left-hand panel shows a deep $HST$ WFPC2 image of the
BR1202--0725 field with faint neighboring continuum structures
circled ($3''$ diameter).  The right-hand panel shows the
corresponding \la\ image taken in a 4-hr integration through an
80\ang-wide narrow-band filter at the f/31 tip-tilt secondary of the
UH 2.2m telescope.  Diffuse \la\ emission was detected coincident
with both NW and SW companion galaxies to the quasar.
\label{fig:br1202}}
\end{center}
\end{figure}

\section*{Acknowledgments}
EMH gratefully acknowledges the support of a University Research Council
Seed Money grant and grant GO-5975 from STScI.  RGM acknowledges the
Royal Society.


\section*{References}

\begin{thebibliography}{99}
\bibitem{rees} M.J. Rees, these proceedings.

\bibitem{lu} L. Lu, \etal,
  \journal{\apjl}{457}{1}{1996}

\bibitem{pritchet} C.J. Pritchet, \journal{\pasp}{106}{1052}{1994}

\bibitem{tdt95} D. Thompson, S. Djorgovski and J. Trauger, 
  \journal{\aj}{110}{963}{1995}

\bibitem{hu96a} E.M. Hu, R.G. McMahon and E. Egami, 
  \journal{\apjl}{459}{L53}{1996}

\bibitem{petit96} P. Petitjean, \etal, \journal{\nat}{380}{411}{1996}

\bibitem{hu96b} E.M. Hu and R.G. McMahon, \journal{\nat}{382}{231}{1996}

\bibitem{egami} E. Egami, PhD thesis, Univ.\ of Hawaii, 1995.

\bibitem{fontana96} A. Fontana, \etal,
  \journal{\mnras}{279}{L27}{1996}

\bibitem{dodorico} S. D'Odorico \etal, these proceedings.

\bibitem{apm} M.J. Irwin, R.G. McMahon and C. Hazard, in {\em The Space 
Distribution of Quasars}, ed. D. Crampton, 117, 1996.

\bibitem{lsl96a} L.J. Storrie-Lombardi, R.G. McMahon, M.J. Irwin and C. Hazard,
 {\apj} in press, [astro-ph/9604021].

\bibitem{apmdla} L.J. Storrie-Lombardi, M.J. Irwin and R.G. McMahon, {\mnras}
 in press, [astro-ph/9608146].

\bibitem{rgm94} R.G. McMahon, \etal,
 \journal{\mnras}{267}{9L}{1994}

\bibitem{isaak} K.G. Isaak, \etal,
 \journal{\mnras}{267}{L28}{1994}

\bibitem{br1202_co} A. Omont \etal, \journal{\nat}{382}{428}{1996}

\bibitem{large_sample} L.L. Cowie, A. Songaila, E.M. Hu and J.G. Cohen, 
 \aj, in press.

\bibitem{gal_form} L.L. Cowie, E.M. Hu and A. Songaila, 
 \journal{\nat}{377}{603}{1995}

\bibitem{charl93} S. Charlot and S.M. Fall, \journal{\apj}{415}{580}{1993}

\bibitem{brock} M. Brocklehurst, \journal{\mnras}{153}{471}{1971}

\bibitem{kenn83} R.C. {Kennicutt, Jr.}, \journal{\apj}{272}{54}{1983}

\end{thebibliography}
\end{document}